\documentstyle[aps]{revtex}
\begin{document}

\title {Moduli-space structure of knots with intersections}
\author {Norbert Grot \thanks{e-mail: norbert@phyast.pitt.edu}
\and Carlo Rovelli \thanks{e-mail: rovelli@pitt.edu} \\
{\it Department of Physics and Astronomy, University of Pittsburgh, 
Pittsburgh, Pa 15260, U.S.A.}}
\date{\today}
\maketitle

\begin{abstract}
It is well known that knots are countable in ordinary knot 
theory.  Recently, knots {\it with intersections} have 
raised a certain interest, and have been
found to have physical applications.  We point out that such knots
--equivalence classes of loops in $R^3$ under diffeomorphisms-- 
are not countable; rather, they exhibit a moduli-space 
structure.  We characterize
these spaces of moduli and study their dimension.  We derive a 
lower bound (which we conjecture being actually attained) on
the dimension of the (non-degenerate components) of the 
moduli spaces, as a function of the valence of the intersection.
\end{abstract}

\section{Introduction}

At the end of his delicious booklet on catastrophe 
theory \cite{arnold}, Arnold notices the following. 
Consider a set of $n$ lines through the origin in the plane. 
Call two such sets equivalent if they can be mapped into each 
other by a linear transformation of the plane. 
The equivalence classes are discrete for $n=1,2,3$; 
but for $n=4$, a moment of reflection shows that the equivalence classes 
are parametrized by a continuous parameter.  Precisely  
this phenomenon is at the root of the emergence of 
a rich moduli space structure in the spaces of knots 
with intersections. 

Knots play an increasingly important role in various areas of mathematics and
physics \cite{Atiyah,RovelliEB,Kauffman}. Classical knot 
theory  \cite{Kauffman2} deals with knots
without intersections, but recent applications of knot theory
require knots {\it with intersections} to be considered as well  
\cite{intersections}.
For instance, quantum states of the gravitational field are 
labeled by knots
with intersections in the loop representation approach to quantum 
gravity
\cite{RovelliSmolin,Outline}. Knots can be defined in two ways: as 
equivalence
classes of loops in $R^3$ under continuous deformations (ambient isotopy) of
the image of the loop -- ``c-knots''; or as equivalence classes (of
unparametrized loops) under invertible smooth transformations (diffeomorphisms)
of $R^3$ -- ``d-knots''. For the non-self intersecting loops, the two
definitions are equivalent and there is no distinction between c-knots and
d-knots.  But the two definitions cease to be equivalent in the case with
intersections.  Intersecting d-knots are different than intersecting
c-knots. The case of intersecting d-knots is of particular interest in physics
\cite{RovelliSmolin}; these knots display a remarkable novel phenomenon, 
which, to our knowledge, has been rarely noticed (the only mention to 
this phenomenon
we could find in the literature is in Ref.\ \cite{baez}): unlike 
ordinary knot spaces, 
the space ${\cal K}_d$ of the intersecting d-knots is not countable.

The continuous dimensions of the space ${\cal K}_d$ come from the differential
structure of the underlying manifold.  
The differential 
structure gives rise to a tangent space $T_p$ at intersection points, 
loops define lines in $T_p$, and diffeomorphisms act linearly on 
$T_p$. Equivalence under diffeomorphisms imply
equivalence under linear trasformations of $T_p$. We are therefore
precisely in the situation of Arnold's example -- one dimension up.
For a large enough number of lines, linear transformations of
$T_p$ fail to be able to align all the lines, and a moduli
space structure emerge. Let us illustrate more in detail
how this comes about
by means of an example. Consider a smooth loop $\alpha$ in $R^3$, with a
self-intersection point $p\in R^3$, and assume that $\alpha$ goes through $p$
five times, so that it has five tangents $\vec v_1,... , \vec v_5$ at
 $p$ (assume any three of the five are
linearly independent).  Let us denote by
$K_c[\alpha]$ the c-knot to which $\alpha$ belongs.  Consider a loop
$\beta$ in the same c-knot $K_c[\alpha]$. The loop $\beta$ will have
an intersection point as well, say $q$, and five tangents $\vec w_1,... , \vec
w_5$ at $q$. In order for $\alpha$ and $\beta$ to be in the same d-knot, there
must be a {\it diffeomorphism} $f: R^3\to R^3$ sending $\alpha$ into
$\beta$. In particular, $f$ maps $p$ to $q$.  
The tangent map $f^*$ maps the tangent space at $p$,
$T_p$, to the tangent space at $q$, $T_q$, and it should align the tangents
$\vec v_i$ ($i=1, ...5$) to the tangents $\vec w_i$.  But $f^*$ is a {\it
linear} map between three-dimensional spaces, given by the Jacobian matrix of
$f$ at $p$; it is a $GL(3)$ transformation depending on 9 parameter.
Since the directions of five vectors $\vec v_i$ depend on 10 parameters, it is
clear that generically no linear transformation exists that aligns five given
vectors $\vec v_i$ to five given vectors $\vec w_i$. Generically $\alpha$ and
$\beta$ will not belong to to the same d-knot.  There will be -at least- one
continuous parameter $\lambda$ --function of the angles between the five
tangents-- which is invariant under diffeomorphisms and distinguishes $\alpha$
from $\beta$. Actually, as we shall see, in this example there are two such
parameters, $\lambda_1$ and $\lambda_2$; we will give them explicitly
below. d-knots are distinguished by such continuous parameters, and therefore
fail to be countable.  The space of all d-knots in $K_c[\alpha]$
is a finite dimensional space obtained by quotienting the infinite
dimensional space $K_c[\alpha]$ by the infinite dimensional group
$Diff_{R^3}$. Namely, it is a moduli space, coordinatized by the two moduli
$\lambda_1$ and $\lambda_2$.  

The same phenomenon repeats in the higher jets --namely for 
derivatives of the loops higher than the tangents-- 
in a more intricate manner. 
Derivatives of order $n$ transform under diffeomorphisms according 
to (non-trivial) 
transformation formulas that depend only on the derivatives 
of order $n-1$ (or lower) of the Jacobian matrix. Since the last
have a finite number of components, 
a sufficiently high number of segments
crossing one intersection will always give rise to new moduli. 
Thus, d-knots are not countable and exhibit a very rich moduli 
space structure, coming from the jets of all orders.

In this paper, we make the above observation precise, we define the moduli
spaces of intersecting d-knots, and study their general structure and their
dimension.  We derive some general results on these dimensions. In particular,
our main result is a formula for the dimension of the (generic components) of
these spaces. We show that this formula gives a lower bound on the dimension of
these spaces as a function of the valence of the intersection, and we
conjecture that the formula gives indeed the correct dimension.  Our original
motivations came from quantum gravity, and we expect, in particular, that our
results could be of interest for that field.

\section{Structure of the intersecting d-knot space}
By loop, 
we indicate here a smooth map $\alpha: S_1\to M$ from the circle $S_1$
to a three-dimensional manifold $M$, which we assume for simplicity having the
topology 
of $R^3$. We indicate loops by Greek letters $\alpha, \beta, ...$, and
denote 
the space of the loops in $R^3$ as $\cal L$.  We consider two
equivalence relations in $\cal L$.  We say that $\alpha$ and $\beta$ are
c-equivalent, and write $\alpha\sim_c\beta$ if there exist a smooth
one-parameter family $c_t, t\in [0,1]$ of smooth, invertible maps from the
image of $\alpha$ to $R^3$ such that $c_0\alpha=\alpha$ and $
c_1\alpha=\beta$. Namely, if the image of $\alpha$ can be smoothly deformed to
the image of $\beta$.  This is clearly an equivalence relation; we call the
corresponding equivalence classes in $\cal L$ c-knots, and denote them as
$K_c$.  We denote the equivalence class to which $\alpha$ belongs as
$K_c[\alpha]$ and the space of c-knots as ${\cal K}_c$. Next we say that
$\alpha$ and $\beta$ are d-equivalent, and we write $\alpha\sim_d\beta$, if
there exist a diffeomorphism $t$ of $S_1$ and a diffeomorphism $f$ of $M$
--connected to the identity-- such that $\alpha=f\circ\beta\circ t$. This too
is an equivalence relation. We call the corresponding d-equivalence classes in
${\cal L}$ d-knots, and denote them as $K_d$. We denote the equivalence class
to which $\alpha$ belongs as $K_d[\alpha]$, and the space of d-knots as ${\cal
K}_d$. Thus
\begin{equation}
        {\cal K}_c={{\cal L}\over{\sim_c}},\ \ \ \ \ 
        {\cal K}_d={{\cal L}\over{\sim_d}}. 
\end{equation}
Our aim is to study the structure of ${\cal K}_d$.  In particular we want to
investigate its continuous dimensions.  d-knots can be labeled by a set of
discrete parameters $k_j$ and continuous parameters $\lambda_j$. We will use a
Dirac-like notation $K_d = |k_j, \lambda_j \rangle$, suggested by the fact that
d-knots label quantum state of spacetime in loop quantum gravity.  We are
interested in studying the appearance and the number of continuous parameters
$\lambda_j$, namely the dimensions of the d-knot moduli spaces.

The space of the c-knots ${\cal K}_c$ is countable. Since every diffeomorphism
in the connected component of the identity induces a smooth deformation of the
image of the loop, $\alpha\sim_d\beta$ implies $\alpha\sim_c\beta$, and
therefore every d-knot is contained inside a single c-knot.  Thus we have a
well defined map $i:{\cal K}_d \to {\cal K}_c$ sending $K_d[\alpha]$ to
$K_c[\alpha]$.  As the example of the introduction shows, the map $i$ is not
injective: a c-knot is formed, in general, by many d-knots.  We call ${\cal
K}_d^{(K_c)}$ the inverse image of $K_c$ under $i$, namely the set of the
d-knots that correspond to the c-knot $K_c$.  The space ${\cal K}_d$ is thus
the union of a countable number of components ${\cal K}_d^{(K_c)}$, one for
every c-knot $K_c$
\begin{equation}
 {\cal K}_d = \bigcup_{K_c\in{\cal K}_c} {\cal K}_d^{(K_c)}.
\end{equation}
In other words, the first discrete parameter that characterizes a d-knot $K_d$
is the c-knot $K_c$ to which it belongs, and we can write  $K_d = |K_c,$ other
parameters$\rangle$.

Let us consider one of the components $K_d^{(K_c)}$. A continuous map cannot
change the number of intersections $I$ of a loop. Therefore this number is well
defined for a c-knot $K_c$.  Each intersection $i$ is further characterized by
the number $N_i$ of times the loop crosses it, which we call the valence of the
intersection, following the literature. Thus a set of integers $N_i, i=1...I$
--the valence of its intersections-- is associated with every c-knot. Imagine
now that three segments cross at the intersection $i$, namely $N_i = 3$.
Imagine that the corresponding three tangents at the intersection are linearly
dependent. A continuous transformation can alter this linear dependence, but a
diffeomorphism cannot. Thus, the presence of linear dependency between tangents
distinguishes d-knot.  We denote an intersection of valence three with linearly
dependent tangents as a degenerate intersection. Similarly, we denote an
intersection of higher valence degenerate if at least one triple of its
tangents is linearly dependent.  As we shall better illustrate below,
degeneracy of this kind --a relation between derivatives of the loop that
cannot be removed by a diffeomorphism-- may happen for higher than first
derivatives of the loops as well. The information about the presence of
degeneracy is discrete, and we represent it collectively by a discrete
parameter $k_i$ for every intersection $i$. We write $K_d = |K_c, k_i, $other
parameters$\rangle$ and denote the set of d-knots in the same c-knot and with
the same degeneracies as ${\cal K}_d^{(K_c, k_i)}$. We shall write $k_i=0$, or
just omit the $k_i$ to indicate that the $i$ intersection has no degeneracies.

This exhausts the discrete parameters that characterize d-knots.  The remaining
parameters distinguishing d-knots are continuous moduli parameters.  Thus, the
space of intersecting d-knots ${\cal K}_d$ can be written as the union of a
denumerable set of components ${\cal K}_d^{(K_c, k_i)}$ as
\begin{equation}
 {\cal K}_d = \bigcup_{K_c\in{\cal K}_c}\  \bigcup_{k_i}\ \  
{\cal K}_d^{(K_c, k_i)}\ ,
\end{equation}
where the spaces ${\cal K}_d^{(K_c, k_i)}$ are finite dimensional moduli
spaces, whose dimensions we are now going to study.

Let us 
consider one of these moduli spaces ${\cal K}_d^{(K_c, k_i)}$.  A moment
of reflection shows that each modulus is attached to one of the intersections,
and that 
there is no relation between moduli of different intersections.  As we
will show 
below, the number of moduli that characterize a d-knot at one
intersection 
depends on the valence $N_i$ of the intersection, and the possible
presence of 
degeneracies described by $k_i$.  Let $d(N_i,k_i)$ be the number of
moduli that 
characterize an intersection $i$.  Then, there will be $d(N_i,k_i)$
continuous 
parameters $\lambda^{(i)}_{j}, j=1...d(N_i,k_i)$ characterizing each
intersection 
$i$. The d-knot is then fully characterized by all these parameter
for each of its intersections.  Namely
\begin{equation}
        K_d = |K_c, k_i, \lambda^{(i)}_{j_i}\rangle, 
\end{equation}
where $i=1...I$  and $j_i=1...d(N_i,k_i)$. In other words, the moduli space
${\cal K}_d^{(K_c, k_i)}$ is 
the cartesian product of one moduli-space per each
intersection. We denote the moduli space of an intersection of valence $N$ and
(possible) degeneracy $k$, by ${\cal K}_{N,k}$. We thus have
\begin{equation}
 {\cal K}_d=\bigcup_{K_c\in{\cal K}_c}\ \bigcup_{k_i}\  
 {\bigotimes}_{i\in K_c}\ \ 
 {\cal K}_{N_i,k_i}.  
\end{equation}
It follows that it is sufficient to study intersections (of any valence $N$
and with any degeneracy $k$) in order to fully determine the general structure
of ${\cal K}_d$.  Below, we will discuss the moduli space ${\cal K}_N={\cal
K}_{N,0}$ of the intersections of arbitrary valence, but with no
degeneracy. The case with degeneracy $k\ne 0$ can be treated along similar
lines.

\section{The moduli space ${\cal K}_N$}

Let $p$ be a non-degenerate intersection point of valence $N$ (we drop the
suffix $i$ since we deal here with a single intersection) in a loop $\alpha$.
We denote by $s$ (or $t,u, ...$) a coordinate on the circle $S_1$ and use
coordinates $x^a$ with $a=1,2,3$ from an atlas of $M$. Thus $p^a$ will be the
coordinates of $p$ and we write $\alpha: s\longmapsto\alpha^a(s)$. There are
$N$ segments of $\alpha$ crossing $p$ (intersection between $\alpha$ and a
sufficiently 
small $M$-neighborhood of $p$); we denote them by $\alpha_i^a(s)$,
where $i=1...N$, and we call $s_i$ the $N$ points in $S_1$ defined by
$\alpha(s_i)=p$. 
Similarly, we consider a second loop $\beta$ in the same moduli
space, namely 
in the same c-knot and with the same degeneracies as
$\alpha$. Let $q$ be its intersection point (corresponding to $p$) and
$\beta_i^a(s)$ the coordinates of the segments crossing in $q$. The two loops
are in the same d-knot if there is a diffeomorphism of the three manifold $f:
x^a\longmapsto f^a(x)$ 
and a diffeomorphism of the circle $t:s\longmapsto t(s)$
such that
\begin{equation}
        f^a(\alpha(t(s))) = \beta^a(s). 
\end{equation}
If we Taylor expand this condition around the intersection point, for each of
the $N$ segments, we obtain
\begin{eqnarray}
\label{expansion}
 \beta^a(s_i) + {\rm d\over{\rm d}s}\beta^a(s)\Big|_{s_i}(s - s_i) + ...
 + {1\over{n!}}{{\rm d}^n\over{\rm d}s^n}\beta^a(s)\Big|_{s_i}(s - s_i)^n
 + ... = f^a(\alpha(t(s_i)))
 \nonumber \\
 + {\rm d\over{\rm ds}}f^a(\alpha(t(s)))\Big|_{s_i}(s - s_i) + ... 
 + {1\over{n!}}{{\rm d^n}\over{\rm ds^n}}f^a(\alpha(t(s)))\Big|_{s_i}
 (s - s_i)^n + ...
\end{eqnarray}
From here on, the 
following notation will be used (for the sake of tradition and brevity):
\begin{eqnarray}
 \dot\alpha^a_i &=& {{\rm d}\over{\rm d}s}\alpha^a(s)\Big|_{s_i},
\nonumber \\
 \alpha^{(n)a}_i &=&
 {{\rm d}^n\over{\rm d}s^n} \alpha^a(s)\Big|_{s_i},
 \nonumber \\
 f^a_{{b_1}...{b_n}} &=&
 \frac{\partial^n f^a}{\partial x^{b_1}...,\partial 
 x^{b_n}}\Big|_p, \nonumber \\
 t^{(n)}_i &=& 
{\rm d^n\over{\rm ds}^n}t(s)\Big|_{s_i}. 
\end{eqnarray}
We now consider each term of the expansion  (\ref{expansion})
 separately.  To zero order, we have  
\begin{equation}\label{zero}
        f^a(p)=q^a. 
\end{equation}
To first order 
\begin{equation}\label{uno}
        f^a_b \dot\alpha^b_i t^{(1)}_i = \dot\beta^a_i.
\end{equation}
Indices are summed if repeated on different levels.  To second order we have 
\begin{equation}\label{due}
 f^a_{bc}\dot\alpha^b_i\dot\alpha^c_i(t^{(1)}_i)^2 +
 f^a_b\dot\alpha^b_it^{(2)}_i = - f^a_b\alpha^{(2)b}_i(t^{(1)}_i)^2 + 
 \beta^{(2)a}_i.
\end{equation}
And for any order $n \geq 2$
\begin{equation}
 f^a_{{b_1}...{b_n}}\dot\alpha^{b_1}_i...\dot\alpha^{b_n}_i(t^{(1)}_i)^n 
 + f^a_b\alpha^{(n)b}_i(t^{(1)}_i)^n + f^a_b\dot\alpha^b_it^{(n)}_i
 + F^a_i = \beta^{(n)a}_i,
\end{equation}
or
\begin{equation}\label{tanti}
 f^a_{{b_1}...{b_n}}\dot\alpha^{b_1}_i...\dot\alpha^{b_n}_i(t^{(1)}_i)^n
 + f^a_b\dot\alpha^b_it^{(n)}_i = \beta^{(n)a}_i -
 f^a_b\alpha^{(n)b}_i(t^{(1)}_i)^n - F^a_i.
\end{equation}
where $F^a_i$ is a function of the derivatives of $f$, $\alpha^a_i$, 
and $t$ of orders lower then $n$ (namely of $f^a_{b_1...b_k}$,...; 
$\alpha^{(k)a}_i$,...;
$t^{(k)}_i$,... with $k=1,...,n-1$).  Equation (\ref{expansion}) is 
equivalent to the infinite system  (\ref{zero}-\ref{tanti}). 

Now, the two loops $\alpha$ and $\beta$ are d-equivalent if this system can be
solved for the functions $f$ and $t$, namely for the infinite tower of
variables $f^a_{b_1...b_n}, t^{(n)}_i$. Therefore, we may regard
(\ref{zero}-\ref{tanti}) as a system of equations for the unknowns
$f^a_{b_1...b_n}, t^{(n)}_i$.  If the system can be solved for every $\alpha$
and $\beta$, then all such loops are in the same d-knot and there are no
moduli. Namely the moduli space has zero dimension. This is the case, for
instance, if $N=2$ (the lowest valence intersection, formed by a single
crossing). In fact, one can check in this case that for each order $n$ the
number of unknowns is larger than the number of equations, and the system can
be solved.  For higher valence intersections, however, the system cannot be
solved for arbitrary $\alpha$ and $\beta$.  This means that there are loops
that are not in the same d-knot, and we have a moduli space structure.

A moment of reflection shows that the dimension of the moduli space is equal to
the number of (independent) equations that overdetermine the system.  To
clarify this point, imagine that the system is solvable for general $\alpha$
and $\beta$ only if we leave, say, $d$ (independent) equations out.  By
inserting 
$f$ and $t$ that solve the rest of the system into these equations we
obtain 
$d$ equations relating $\alpha$ and $\beta$. If we imagine that $\beta$
is fixed, we obtain then $d$ conditions on $\alpha$, determining the set of
$\alpha$'s d-equivalent to $\beta$. Thus, this set has codimension $d$ in the
space of the $\alpha$'s. This means that a d-knot has codimension $d$ in the
space of the loops in ${\cal K}_N$, and therefore that there is a $d$
parameters space of d-knots in ${\cal K}_N$. Namely ${\cal K}_N$ is
$d$-dimensional.

Our task is therefore to find --for every given $N$-- the number $d$ of
independent equations by which the system (\ref{zero}-\ref{tanti}) is
overdetermined. 
This may seem a hard task, given that the system has an infinite
number of 
equations, but there is a key observation that simplifies the matter.
First 
observe that the system has a rather simple structure. As we increase the
order $n$, 
at each new order there are only a finite number of new unknowns that
appear. 
Indeed the unknowns $f^a_{b_1...b_n}$ and $t^{(n)}_i$ appear only at
order $n$ or 
higher.  We denote them as unknowns of order $n$. For instance, at
order 
zero, the only unknowns are the three $f^a$. At order one, we have the
new unknowns $f^a_b$ (nine of them) and $t^{(1)}_i$ (N of them), and so
on. Now, at each order $n$, we have the same number 3$N$ of equations in the
system.  But the number of unknowns increases rapidly, because the number of
components of $f^a_{b_1...b_n}$ increases with $n$. Indeed,
$f^a_{{b_1}...{b_n}}$ has $3\times\rm I_n$ independent entries, where
\begin{equation}
 I_n = {{(n + 1)(n + 2)}\over2}
\end{equation} 
is the number of independent components in a completely symmetrized
$3\times3\times... \times 3 $ $n$-dimensional matrix.  It is then easy to see
that (for fixed $N$) the equations of sufficiently high order can always be
solved.  More precisely, for every $N$, there is a number $m$, which we
determine 
below, such that all equations of order higher than $m$ can always be
solved, and we can safely forget them.  This fact essentially reduces the
system to a finite dimensional system, making the problem treatable.

\subsection{A gauge}

One is now 
tempted to immediately proceed to determine the number of equations
by which the system is overdetermined by naively counting equations and
unknowns order by order, and subtracting.  Unfortunately, there is a
complication.  
At every order $n$, the actual number of unknowns is less than
what a 
simple count would suggest, because of the particular structure of our
equations. Consider first equation (\ref{uno}). The unknowns are the 9
components of $f^a_b$ and the $N$ quantities 
$t^{(1)}_i$.  However, if $f^a_b, t^{(1)}_i$
solve (\ref{uno}), so do
\begin{equation} \label{gauge1}
        \tilde f^a_b = T f^a_b,\ \ \ \ \ \tilde t^{(1)}_i= T^{-1} t^{(1)}_i
\end{equation}
for every non vanishing $T$.  Therefore, the overall scale $T$ can never be
determined by equation (\ref{uno}). In other words, equation (\ref{uno})
depends on only $(9+N-1)$ functions of the $(9+N)$ quantities $f^a_b,
t^{(1)}_i$.  The remaining one cannot be determined by this equation.

The same happens at higher orders. It is easy to verify that if
$f^a_{b_1...b_n}, t_i^{(n)}$ solve the equation of order $n$, so do
\begin{eqnarray}
 \tilde f^a_{b_1..b_n} &=& f^a_{b_1..b_n} + f^a_{(b_1}T_{b_2...b_n)}, 
 \nonumber \\
 \tilde t^{(n)}_i &=& t^{(n)}_i -
 T_{a_1..a_{n-1}}\dot\alpha^{a_1}_i..\dot\alpha^{a_{n-1}}_i (t^{(1)}_i)^2.
\end{eqnarray}
for every symmetric tensor $T_{a_1..a_{n-1}}$.  This tensor has $I_{n-1}$
components.  We call this transformation the n-order gauge of the system.
Because of the gauge, if we cut off the system at order $n$, we have indeed
$I_{n-1}$ less unknowns entering the system than what a naive counting would
suggest.  
Are there other degeneracies beside the gauge we have just described?
We suspect there aren't, but we have not been able to prove this in
general. Because of this incompleteness, we cannot claim that the number we
compute 
below is in fact the dimension of the moduli space, but only that it is
the dimension's lower bound.

\subsection{Size of the space of solutions}

Consider the order $n=0$, 
equation (\ref{zero}). We have three unknowns ($f^a$)
and three equations there. The system is linear and can obviously always be
solved. Consider next the order $n=1$, equation (\ref{uno}). There are $9+N$
unknowns, and $3N$ equations. But, because of the gauge described above, only
$9+N-1$ unknowns can be determined by the equations. Generically, the system
can be solved if the the number of equations is less than or equal to the
number of unknowns, namely if
\begin{equation}
 3 N \le 9 + N -1. 
\end{equation}
In this case, if $N\le 4$.  A simple inspection of the equation confirms that
for $N\le 4$ the 3 equation can indeed be solved, and thus they do not give
rise to any 
continuous dimension. (Since we assumed absence of degeneracies at
the beginning of our analysis.) What happens if $N=5$?  In this case we have
15 equations and 13 (independent) unknowns.  Which means that the system is
overdetermined by two equations. Again, inspection shows that this is 
indeed the
case. Correspondingly, we expect to have at least a two-dimensional moduli
space for $N=5$.

Let us study this $N=5$ case.  At order 2 we have $3N=15$ equations and
$3I_2+N-I_{2-1}+I_{1-1}=21$ unknowns, where the term $I_{2-1}$ 
represents the
number of irrelevant variables (the ones that cannot be solved for) because 
of
the gauge of order 2, and the term $I_{1-1}$ represents the gauge unknown of
order 1 that becomes relevant (can be solved for) at order 2.  We have more
unknowns than equations and so we expect the system to be solvable. 
Indeed, it is solvable. 
The same happens for higher orders, and thus we can conclude
that the only equations for which the system is overdetermined are 
the two of
order 1. Thus, an intersection of valence $5$ has a two dimensional moduli
space. In the next section, we will study this example in detail for
illustration. Here let us continue the general analysis.

At any given order $n$, we have $3N$ equations and 
$3I_n+N-I_{n-1}+I_{n- 2}$
new unknowns where, again the term $I_{n-1}$ is for the gauge terms of
 order
$n$ (that are not being solved for at order $n$) and the 
term $I_{n-2}$ is for
the gauge terms of order $n-1$ 
(that can be solved for in order $n$ as opposed
to order $n-1$). Generically, the system can be solved if the number of
equations is less than the number of unknowns:
\begin{equation}
 3N \le 3 I_n + N - I_{n-1} +  I_{n-2},
\end{equation}
which yields
\begin{equation}\label{ncond}
 N \leq {3n^2 + 7n + 6\over 4}.
\end{equation}
Solving this for $n$, we find that the system is solvable 
at any order $n>m(N)$ where
\begin{equation}\label{mofN}
m(N) =: {\rm Int}^- \Big({\sqrt{48 N - 23} - 7 \over 6} \Big)
\end{equation}
where ${\rm Int}^-(x)$ is the largest integer smaller than $x$.  Thus we
can forget all equations of order larger than $m(N)$ in the system
((\ref{zero})-(\ref{tanti})).  The remaining system is formed by 
the $3N\times
m$ equations of order $n\le m(N)$. (We do not count the 3 equations of order
zero and the 3 unknowns $f^a$, which can be always found.) At each order $n$
(less or equal to $m$), the number $d_n$ of overdetermined equations is
\begin{equation}
 d_n=\# equations-\#unknowns =  3N - (3I_n + N - I_{n-1}+I_{n-2}),
\end{equation}
(where $I_0=I_{-1}=0$), and the total number of equations by 
which the system is overdetermined is
\begin{equation}
 d=\sum_{n=1,m} d_n =\sum_{n=1,m} 2N - 3I_n + I_{n-1}-I_{n-2}
= 2mN + I_{n-1}
 -3 \sum_{n=1,m}I_n.
\end{equation}
Since we are under the assumption that the intersection
is nondegenerate, there are no additional degeneracies 
in the linear system, and the $d$ equations by which the
system is overdermined are independent.  Thus $d$ 
gives the lower bound
on the dimension of the moduli space that we are searching. 
Performing the sum, we get 
\begin{equation}\label{mainres}
 d(N) = \Big(2 N - 5 \Big)m - {5\over2}m^2 - {1\over2}m^3,
\end{equation}
where $m$ is a function of $N$, given in Eq.\ (\ref{mofN}).  Equation 
(\ref{mainres}) is our main result. It gives (a lower bound on) the 
dimension of the moduli space of a single nondegenerate 
intersection of order $N$.

\section{An example: $N$ = 5}

Let's consider again the simplest non-degenerate case in
which a moduli-space appear, which is $N = 5$. Thus, we
have an intersection point $p$ crossed by the loop $\alpha$ 
five times. 
From Eq.\ (\ref{mofN}) we have $m= 1$ and from Eq.\ (\ref{mainres}) 
 $d=2$, as anticipated.   
It is instructive to identify the two continuous degrees
of freedom of the knot space.  We will do this in two 
different ways. First we give a geometrical construction of these
two degrees of freedom, and then we give an explicit algebraic
expression for the two moduli.

Let us fix an arbitrary coordinate chart in the neighborhood
of $p$, and let $\dot\alpha^a_i$ for $i=1,2,3,4,5$ be the components
of the five tangents of $\alpha$ at $p$.  Let us (arbitrarily)
pick three of these these five tangents, say $\dot\alpha^a_k$ 
for $k=1,2,3$.  The three vectors $\dot\alpha^a_k$ define a basis
in the tangent space at $p$. Clearly the components of the other
two tangents, $\dot\alpha^a_4$ and $\dot\alpha^a_5$ on this
basis are quantities that do not depend on the coordinate chosen, 
and are consequently invariant under diffeomorphisms. 
If we indicate by $(\dot\alpha^{-1})^k_a$ the 3x3 matrix 
inverse to the 3x3 matrix 
$\dot\alpha^a_k$, such components are given by
\begin{eqnarray}
	\beta_4^k &=& (\dot\alpha^{_1})^k_a\dot\alpha^a_4,
\nonumber \\ 
	\beta_5^k &=& (\dot\alpha^{_1})^k_a\dot\alpha^a_5.
\end{eqnarray}
The quantities $\beta_4^k$ and $\beta_5^k$ are invariant under
diffeomorphisms.  They transform under a reparametrization of 
the loop as $\beta_4^k\longmapsto t_4 t^{-1}_k \beta_4^k$, 
where
$t_i$ are the 5 derivative of the reparametrization in the
intersection point.  Assuming, for instance, that the components of 
$\beta_4^k$ are positive, we can always choose these derivatives 
in such a way that, say, $\beta_4^k = (1,1,1)$. The length of
the last vector, $\beta^k_5$, can be arbitrarily rescaled, by
fixing $t_5$, 
but its direction is uniquely determined. This direction gives the
two dimensions of the moduli space.  

Notice that the sign
of the components of  $\beta_4^k$ determine eight 
disconnected 
sectors of the moduli space, at the boundary of which are 
degenerate intersections. This is a general feature: the 
moduli spaces in general have disconnected components, separated 
by the degenerate cases. 

Given the above discussion, it is not too hard to 
write the two moduli explicitly. This can be done, 
for instance, in the
following manner
\begin{equation}
 \lambda_1=
 {(\dot\alpha^{-1})^1_a\dot\alpha^a_4\  
(\dot\alpha^{-1})^2_b\dot\alpha^b_5\over
 (\dot\alpha^{-1})^2_c \dot\alpha^c_4  \ 
(\dot\alpha^{-1})^1_d \dot\alpha^d_5},\ \ \ \ \ \ 
 \lambda_2=
 {(\dot\alpha^{-1})^1_a\dot\alpha^a_4 \ 
(\dot\alpha^{-1})^3_b\dot\alpha^b_5 \over
 (\dot\alpha^{-1})^3_c \dot\alpha^c_4 \ 
(\dot\alpha^{-1})^1_d \dot\alpha^d_5}.
\end{equation}
It is easy to see that these two quantities are independent
and are invariant under diffeomorphisms of the manifold and reparametrization
of the loops. 

\vskip2cm

We thank Ted Newman for some suggestions. We are particularly 
indebited to Lee Smolin: the problem considered here emerged 
in a discussion with him.  
This work was partially supported by NFS Grant 
5-39634.

\end{document}